\documentclass[twocolumn,pra,amssymb]{revtex4}

\def\th{^{\mbox{\scriptsize th}}}

\newcommand{\ket}[1]{\mbox{$| #1 \rangle$}}
\newcommand{\bra}[1]{\mbox{$\langle #1 |$}}

\def\H{\mathcal{H}}
\def\J{\mathcal{J}}
\def\K{\mathcal{K}}
\def\x{{\bf x}}
\def\inv{^{\mbox{\scriptsize inv}}}
\def\sol{^{\mbox{\scriptsize sol}}}

\def\t{^{\mbox{\tiny T}}}
\def\c{\mathcal{C}}
\def\tr{\mbox{tr}}
\def\o{\!\otimes\!}
\def\id{\mathbb{I}}

\newcommand{\CC}{{\mathbb{C}}}
\def\vs{\vspace{.2cm}}

\begin{document}

\title{Useful entanglement can be extracted from all nonseparable states}
\author{Llu\'{\i}s Masanes}
\affiliation{School of Mathematics, University of Bristol, Bristol
BS8 1TW, U.K.}
\date{\today}

\begin{abstract}
We consider entanglement distillation from a single-copy of a
multipartite state, and instead of rates we analyze the
``quality'' of the distilled entanglement. This ``quality'' is
quantified by the fidelity with the GHZ-state. We show that each
not fully-separable state $\sigma$ can increase the ``quality'' of
the entanglement distilled from other states, no matter how weakly
entangled is $\sigma$. We also generalize this to the case where
the goal is distilling states different than the GHZ. These results provide new insights on the geometry of the set of separable states and its dual (the set of entanglement witnesses).
\end{abstract}

\maketitle

\section{Introduction}

Quantum Information Science studies the possibilities and
advantages in using quantum physical devices for information
processing tasks. Some of these tasks require noiseless
entanglement as a principal ingredient \cite{NC}. Real devices are
noisy and not isolated from the environment, therefore they
usually cannot be described by pure states. Entanglement
distillation is the process of transforming noisy into noiseless
entanglement by using only {\em local operations and classical
communication} (LOCC) \cite{distillation}. In the most studied
scenario, one has an arbitrarily large number of copies of a state
and wants to know at which rate singlets can be obtained. {\em
Bound entangled states} are the ones that cannot be distilled, and its
existence is proven in \cite{bound-e}. \vs

In this paper we generalize the results of \cite{m} to states with
an arbitrary number of parties. However, as explained in what follows, this generalization is made in the strongest possible way. In the multipartite scenario the
entanglement present in a particular state is not necessarily
shared among all parties. Here we call entangled any state $\rho$
that is not fully-separable, that is
\begin{equation}\label{fully-sep}
    \rho \neq \sum_k \varrho_1^k\otimes\cdots\otimes\varrho_N^k\ ,
\end{equation}
where $\varrho_n^k$ are positive semi-definite matrices. We show that all entangled (i.e. not fully-separable) states have some extractable GHZ-like
entanglement \cite{ghz}. This provides a very simple picture of
quantum correlations, in which all kinds of entanglement
(independent of the number of parties they involve) are, in a sense,
fungible.\vs


\section{Single-copy distillation scenario}

Given an
arbitrary $N$-partite state $\rho$ (acting on $\H=\bigotimes_n^N
\H_n$) we consider the states $\tilde{\rho}$ acting on
$\K=\bigotimes_n^N \CC^2$ that can be obtained from $\rho$ by LOCC
with some probability. This probability can be arbitrarily small
as long as it is nonzero. This class of transformations is called
stochastic-LOCC (SLOCC). Each of these states $\tilde{\rho}$ is
the normalized output of a separable (not necessarily
trace-preserving) completely-positive map \cite{NC}, with $\rho$ as
input:
\begin{equation}\label{map}
    \tilde{\rho} = \frac{\Omega( \rho)}
    {\tr\, \Omega(\rho)}\ .
\end{equation}
By separable map we mean one that can be written as
\begin{equation}\label{pp}
   \Omega(\rho)= \sum_k \left[ M_1^{k}\o\cdots\o
   M_N^{k}\right]
   \rho \left[ M_1^{k}\o\cdots\o M_N^{k} \right]^\dag\ ,
\end{equation}
with $M_n^{k}:\H_n\rightarrow \CC^2$ for all $k$ and $n=1\ldots
N$. We consider SLOCC transformations because we do not care about
the rates at which the states $\tilde{\rho}$ can be obtained from
$\rho$. Instead, we want to know which of the states
$\tilde{\rho}$ resembles more to the GHZ-state. We quantify this
resemblance by the fidelity that a state $\tilde{\rho}$ has with
the $N$-partite GHZ state
\begin{equation}\label{ghz}
    \ket{\Phi}=\frac{1}{\sqrt{2}}
    \big(\ket{0\cdots 0}+\ket{1\cdots 1}\big)\ .
\end{equation}
Define $E(\rho)$ as the largest overlap with the GHZ that a
state $\tilde{\rho}$ obtainable from $\rho$ by SLOCC can achieve:
\begin{equation}\label{mesf}
    E (\rho)=\sup_{\Omega \in \mbox{\scriptsize SEP}}
    \frac{\tr\left[\Omega(\rho)\, \Phi \right]}
    {\tr\, \Omega(\rho)}\ ,
\end{equation}
where $\Phi$ is the projector onto the $N$-partite GHZ state
(\ref{ghz}), and the supremum is taken over all maps of the form
(\ref{pp}) for which $\tr\left[ \Omega(\rho)\right]>0$. \vs

The reason for writing $E(\rho)$ as a supremum instead of a
maximum is because, for some states $\rho$, the set of numbers
$\{\tr[\tilde{\rho}\, \Phi]$ : $\tilde{\rho}$ obtainable from
$\rho$ by SLOCC$\}$ does not have a maximum. In such cases, the
probability of obtaining $\tilde{\rho}$ from $\rho$ goes to zero
as $\tr[\tilde{\rho}\, \Phi]$ goes to $E(\rho)$. When $E(\rho)=1$
this phenomenon is called quasi-distillation
\cite{activation,mesf}. In \cite{dc} it is shown that if
$E(\rho)>1/2$ one can asymptotically distill $N$-partite
GHZ-states from $\rho$. In complement, $E(\rho)\geq 1/2$ holds for
any $\rho$, because the state $\ket{0\cdots0}$ can be prepared
locally and its fidelity with $\Phi$ is $1/2$. Therefore, the range
of $E$ is $[1/2,1]$. \vs

By definition (\ref{mesf}), the quantity $E(\rho)$ is
nonincreasing under SLOCC processing of $\rho$, and thus, an
entanglement monotone \cite{Vidal}. It is also an operationally meaningful
entanglement measure in the context of single-copy distillation:
$E(\rho)$ is the probability that a state distilled from $\rho$
``looks'' like the GHZ state. In the bipartite case ($N=2$),
$E(\rho)$ is related to the average fidelity of the conclusive
teleportation channel obtainable from $\rho$, denoted $F(\rho)$.
In \cite{mesf} it is shown that
\begin{equation}\label{}
    F(\rho)=\frac{2 E(\rho) +1}{3}\ .
\end{equation}
The quantity $E$ also allows us to express our main result in
a compact way. \vs

\section{Results}

{\bf Theorem 1.} {\em An $N$-partite state $\sigma$ is entangled
if, and only if, for any $\lambda\in[1/2,1)$ there exists an
$N$-partite state $\rho$ such that $E(\rho)\leq \lambda$ and
$E(\rho\otimes\sigma) >\lambda\,$.} \vs

In other words, chose a threshold $\lambda$ on the ``quality'' of
the distilled entanglement above which you are satisfied, and
consider the set of states $\rho$ from which it is impossible to
distill such a sufficiently good entanglement $E(\rho) \leq \lambda$. Any entangled state $\sigma$ (no matter how weakly entangled it is) can increase the ``quality'' of
the entanglement distilled from some states $\rho$ (in the above
mentioned set) to a value larger than the threshold $\lambda$.
In particular, if we choose a high threshold (e.g. $E(\rho) \leq 0.999$), any entangled state $\sigma$ contributes by producing a result with even  more purity ($E(\rho\otimes\sigma) > 0.999$). Then, one can argue that
GHZ-like entanglement is being extracted form
$\sigma$. Also remarkably, notice that $\sigma$ could factorize with respect to some parties (or equivalently, $\sigma$ is only shared by less than $N$ parties), and yet, $\sigma$ enhances the full $N$-partite GHZ-like entanglement present in other states! Hence, the way Theorem 1 generalizes the results proven in \cite{m} is, unexpectedly, the strongest possible one. In what follows, we provide another surprising example. \vs

Let us consider a three-qubit state $\sigma_{\mbox{\scriptsize shifts}}$, presented in \cite{shifts}, which has remarkable properties. The state  $\sigma_{\mbox{\scriptsize shifts}}$ is proportional to the projector onto the subspace orthogonal to the unextendible product basis $\{\ket{0,1,+}, \ket{1,+,0}, \ket{+,0,1}, \ket{-,-,-}\}$, where $\{\ket{0}, \ket{1}\}$ is an orthonormal basis and $\ket{\pm} =\ket{0}\pm \ket{1}$. This state is not fully-separable (\ref{fully-sep}), but when  any two of the three parties are considered as a single one, the resulting bipartite state becomes separable. This property may suggest that $\sigma_{\mbox{\scriptsize shifts}}$ does not contain useful quantum correlations. Yet, according to Theorem 1, 3-partite GHZ-like entanglement can be extracted from $\sigma_{\mbox{\scriptsize shifts}}$, but also 4-partite GHZ-like entanglement, and so on.\vs

One can generalize Theorem 1 to more intricate
single-copy distillation scenarios. We denote by $S$ any subset of
more than one party $S\subseteq\{1,\ldots N\}$, and by
$\ket{\Phi_S}$ the $|S|$-partite GHZ-state (\ref{ghz}) shared among all parties
in $S$. Consider $M$ disjoint subsets of this kind
$S_1,\ldots S_M$ and a subset $R$ containing the rest of parties:
$R= \{1,\ldots N\}\backslash\bigcup_{m=1}^M S_m$. Suppose that the
parties within each subset $S_1,\ldots S_M$ aim at distilling a
shared GHZ state, and the rest of parties $R$ help to achieve this
goal. This scenario motivates the definition of a quantity $Q$
which generalizes E:
\begin{equation}\label{Q}
    Q_{S_1\cdots S_M}(\rho)=
    \sup_{\Omega \in \mbox{\scriptsize SEP}}
    \frac{\tr\left[\Omega(\rho)\,
    \id_{R}\!\otimes\Phi_{S_1}\otimes\cdots\otimes\Phi_{S_M}\right]}
    {\tr\, \Omega(\rho)}\ .
\end{equation}
Clearly, $Q_{\{1,\ldots N\}} =E$. One can check that the range of $Q_{S_1\cdots S_M}$ is
$[2^{-M},1]$. Theorem 1 can be generalized to this
more intricate distillation scenario.\vs

{\bf Theorem 2.} {\em An $N$-partite state $\sigma$ is entangled
if, and only if, for any partition of the set of parties
$S_1,\ldots S_M,R$ and any $\lambda\in[2^{-M},1)$ there exists an
$N$-partite state $\rho$ such that $Q_{S_1\cdots S_M}(\rho)\leq
\lambda$ and $Q_{S_1\cdots S_M}(\rho\otimes\sigma)
>\lambda\,$.} \vs

This result has a similar interpretation than Theorem 1, but its
consequences are more rich. Imagine a 5-partite scenario where parties $\{1, 2\}$ want to distill a singlet $\ket{\Phi_{\{1,2\}}}$, or any state whose fidelity with
$\ket{\Phi_{\{1,2\}}}$ is larger than $0.9$. Parties $\{3,4,5\}$
share the supposedly useless state $\sigma_{\mbox{\scriptsize shifts} \{3,4,5\}}$. Theorem 2 demonstrates the existence of a 5-partite state $\rho_{\{1,\ldots 5\}}$ from which it is
impossible to extract any state whose fidelity with
$\ket{\Phi_{\{1,2\}}}$ is larger than $0.9$, but, together with
$\sigma_{\mbox{\scriptsize shifts} \{3,4,5\}}$ this goal can be achieved. Notice that $\sigma_{\mbox{\scriptsize shifts} \{3,4,5\}}$ does not
involve parties $\{1, 2\}$! Therefore, in the joint ``activation" of $\rho_{\{1,\ldots 5\}}$ and $\sigma_{\mbox{\scriptsize shifts} \{3,4,5\}}$, some intrincate teleportation-like phenomena between the sets of parties $\{1,2\}$ and $\{3,4,5\}$ take place.
\vs

\section{Proofs}

In this section the core of the proofs of Lemmas 1 and 2 is included. The remaining part of the proofs can be found in the appendix.\vs

{\em Proof of Theorem 1.} If $\sigma$ is fully-separable then $E(\rho\otimes\sigma)=E(\rho)$
for any $\rho$. This holds because fully-separable states can be
created by LOCC, and the definition of $E$ already involves an optimization over LOCC. Let us prove the other direction of the equivalence.

From now on $\sigma$ is an arbitrary $N$-partite not
fully-separable state acting on $\H=\H_1\otimes\cdots\otimes\H_N$,
and $\lambda$ is fixed to some arbitrary value within $[1/2,1)$.
We have to show that there always exists a state $\rho$ such that
$E(\rho)\leq \lambda$ and $E(\rho\otimes\sigma)>\lambda$. We fix
$\rho$ to be an $N$-partite state for which the $n\th$ party's
Hilbert space is $\J_n\!\otimes\K_n$, where $\J_n=\H_n$ and
$\K_n=\CC^2$, for $n=1\ldots N$. We also define
$\J=\bigotimes_{n}^N \J_n$ and $\K=\bigotimes_n^N \K_n$.

Given a finite list of pairs of positive numbers $(x_1,y_1),\ldots
(x_n,y_n)$ the following inequality can be proven by induction:
\begin{equation}\label{ineq}
    \frac{x_1+\cdots+x_n}{y_1+\cdots+y_n}\, \leq\,
    \max_k \frac{x_k}{y_k}\ .
\end{equation}
Using it, one can see that the the supremum in expression
(\ref{mesf}) is always achievable by a map $\Omega$ with only one
term:
\begin{equation}\label{E}
    E(\rho)=\sup_{{\bf M}}
    \frac{\tr\left[ {\bf M}\, \rho\,
    {\bf M}^\dag \Phi \right]}
    {\tr\left[ {\bf M}\, \rho\,
    {\bf M}^\dag \right]}\ ,
\end{equation}
where ${\bf M}$ is any product matrix ${\bf M}= M_1\o\cdots\o
M_N$ with $M_n:[\H_n\otimes\CC^2] \rightarrow \CC^2$ for
$n=1\cdots N$. Using (\ref{E}) one can characterize the set of
states $\rho$ satisfying $E(\rho)\leq \lambda$, by the following
set of linear inequalities:
\begin{equation}\label{set}
    \tr\!\left[ \rho\ {\bf M}^\dag
    \left(\lambda \mathbb{I} -\Phi\right)
    {\bf M} \right] \geq 0\quad
    \forall\ {\bf M}=M_1\o\cdots\o M_N
\end{equation}
where $\mathbb{I}$ is the identity matrix acting on $\K$. For
convenience, in the rest of the proof $\rho$ is allowed to be not
normalized. We denote by $\c$ the set of states satisfying all the
inequalities (\ref{set}):
\begin{equation}\label{}
    \c=\{\rho : \rho\geq 0,\, E(\rho)\leq \lambda\}\ ,
\end{equation}
which is a convex cone. The dual cone of $\c$ is
\begin{equation}\label{dual}
    \c^*=\{X : \tr[\rho\, X]\geq 0\
    \forall \rho\in\c\}\ .
\end{equation}
A generalized version of Farkas Lemma \cite{farkas} states that any matrix
$X\in\c^*$ can be written as
\begin{equation}\label{element}
    X=\sum_k {\bf M}_k^\dag
    \left(\lambda \mathbb{I} -\Phi\right){\bf M}_k\ +
    \sum_s P_s\ ,
\end{equation}
where $P_s$ are positive matrices, and ${\bf M}_k$ are arbitrary product matrices like in (\ref{set}).

Let us concentrate on the condition
$E(\rho\otimes\sigma)>\lambda$. Instead of computing the supremum
in (\ref{E}) we consider a particular filtering operation
$\tilde{{\bf M}}$, with which we obtain a lower bound on
$E(\rho\otimes\sigma)$. The chosen form of $\tilde{{\bf M}}$ is
\begin{equation}\label{tildeM}
  \tilde{M_n} = \bra{\phi_{\H_n\J_n}}\otimes
  \mathbb{I}_{\K_n}\quad n=1\ldots N\ ,
\end{equation}
where $\ket{\phi_{\H_n \J_n}}$ is the (local) maximally entangled
state between the systems corresponding to $\H_n$ and $\J_n$
(which have the same dimension), and $\mathbb{I}_{\K_n}$ is the
identity matrix acting on $\K_n$. A little calculation shows that
for any matrix $Z$ acting on $\K$, the equality
\begin{equation}\label{eq}
    \!\!\tr\!\left[ \tilde{\bf M}\, (\rho_{\J\K} \otimes
    \sigma_{\H})\, \tilde{\bf M}^\dag\, Z \right]
    =\nu\,
    \tr \left[\rho_{\J\K}\, (\sigma_{\J}\t \otimes
    Z_{\K}) \right]
\end{equation}
holds, where $\sigma\t$ stands for the transpose of $\sigma$, and $\nu >0$. In
the above expression the sub-indexes $\H,\J,\K$ explicitly indicate
on which Hilbert spaces every matrix acts, and with which other
matrices its indexes are contracted. Using (\ref{eq}), a
sufficient condition for $E(\rho\otimes\sigma)>\lambda$ is
\begin{equation}\label{cond}
    \tr\left[\rho \left(\sigma\t
    \o(\lambda\mathbb{I}-\Phi)\right) \right]
    <0\ .
\end{equation}
Let us show that there always exists a $\rho\in\c$ satisfying this
inequality, by creating a contradiction.

Suppose that no single $\rho\in\c$ satisfies (\ref{cond}). Then,
by definition (\ref{dual}), the matrix $\sigma\t \otimes
(\lambda\mathbb{I}-\Phi)$ belongs to $\c^*$, and we can express it
as in (\ref{element}). One way of writing this is
\begin{equation}\label{principal}
    \sigma\t \otimes (\lambda\id-\Phi) -
    \Omega\! \left(\lambda \id -\Phi\right)
    \geq 0\ ,
\end{equation}
where $\Omega$ maps matrices acting on
$\K$ to matrices acting on $\H\o\K$, and is separable (\ref{pp}). In the rest of the proof we
also use the same symbol $\Omega$ to denote any separable
completely-positive map. This is not confusing because each of
these maps is arbitrary, and the input and output spaces of each
$\Omega$ is unambiguously fixed by the context. Performing the
partial trace over $\H$  to (\ref{principal}) we obtain
\begin{equation}\label{uk}
    (\lambda\id-\Phi) -
    \Omega\! \left(\lambda \id -\Phi\right)
    \geq 0\ .
\end{equation}
Notice that here $\Omega$ is different than in (\ref{principal}), but still separable.
In \cite{dc} it was presented a depolarization protocol (here
denoted by $\Delta$) that can be implemented by LOCC, and leaves
invariant the identity and the GHZ state: $\Delta(\id)=\id$ and
$\Delta(\Phi)=\Phi$. Because $\Delta$ is a completely positive
map, we can apply it to the left-hand side of (\ref{uk}) obtaining
the positive matrix
\begin{equation}\label{ug}
    (\lambda\id-\Phi) -
    [\Delta\circ\Omega\circ\Delta]\! \left(\lambda \id -\Phi\right)
    \geq 0\ .
\end{equation}
Lemma 1 in the Appendix shows that such an $\Omega$ must fulfill $(\lambda\id-\Phi) - [\Delta\circ\Omega\circ\Delta]
(\lambda\id-\Phi) =0$. Hence, the trace of the left-hand side of (\ref{principal}) is zero, and a positive traceless matrix can only be the null matrix. Using the
properties of $\Delta$ we can write
\begin{equation}\label{ku}
    \left[(\id_{\H}\otimes\Delta_{\K})\circ\Omega\circ\Delta\right]\!
    \left(\lambda \id -\Phi\right)
    =\sigma\t \otimes(\lambda\id-\Phi)\ .
\end{equation}
Lemma 2 in the Appendix tells us that $\sigma\t$ must be separable. But this
is in contradiction with the initial assumption that $\sigma$ is
entangled. Therefore, the supposition that no single $\rho\in\c$
satisfies (\ref{cond}) is false. $\Box$ \vs

{\em Proof of Theorem 2.} Here one proceeds in the same way as
before, but instead, the depolarization protocol $\Delta$ is applied
independently by each subset of parties $S_m$. In other words, one has to proceed with the global depolarization map $\Delta_{S_1}\otimes\cdots
\otimes\Delta_{S_M}\otimes \id_R$. $\Box$\vs

\section{Final remarks}

The statement of Theorem 1 can be made
stronger for the case $\lambda=1/2$ . The SLOCC operation
(\ref{tildeM}) that transforms $\rho\otimes\sigma$ into a state
having fidelity with the GHZ larger than $1/2$, can be substituted
by a deterministic operation (LOCC). This is done by using the
following trick. When the SLOCC
operation (\ref{tildeM}) succeeds, the $N$ parties do nothing, and
when it fails they substitute the residual state by $\ket{0\cdots
0}$, whose fidelity with the GHZ is 1/2. Clearly, the mixture of the success and failure states has
fidelity with the GHZ strictly larger than $1/2$. Therefore, all that we have said is not a particular fact of SLOCC transformations, everything can be done with probability one.

\vs

The method used to prove these theorems is non-constructive, hence it does not say much about
the state $\rho$, whose entanglement is enhanced by $\sigma$. The
only thing we know about $\rho$ is that it is related to an
entanglement witness \cite{KLC} that detects $\sigma$. Precisely,
the operator
\begin{equation}\label{witness}
    W_{\H}=\tr_{\K}\left[ \rho_{\H\K}
    \, (\lambda\mathbb{I}_{\K}-\Phi_{\K}) \right]\ ,
\end{equation}
can be proved to be an entanglement witness by imposing
$E(\rho)\leq \lambda$. That $W$ detects $\sigma$ follows from
inequality (\ref{cond}). As a consequence of Theorem 1, the set of
witnesses of the form (\ref{witness}) is complete, in the sense
that they detect all entangled states for any number of parties.
\vs

It is straightforward to generalize the depolarization protocol in
\cite{dc} to larger local dimension ($d>2$). This suggests that
these theorems also hold when considering the overlap with the
higher dimensional state
\begin{equation}\label{}
    \ket{\Phi}= \frac{1}{\sqrt{d}}
    \sum_{k=1}^d \ket{k\cdots k}\ .
\end{equation}
It would be interesting to know if Theorem 1 can be generalized
when the fidelity is measured with respect to an arbitrary
entangled pure-state, or contrary, there are states for which such
a theorem does not hold. That would provide deep insights on the structure of multipartite entanglement.\vs

\section{Conclusions}

We have shown that all entangled states can increase
the ``quality'' of the entanglement distillable from a
single-copy of other states. Obviously this task is impossible for
fully-separable states. Hence, all quantum correlations have a
qualitatively different character. Then we can say more than ``entanglement is a
physical resource"; we can say that ``{\em all} entanglement is a physical
resource"!

\section{Acknowledgments}

The author is thankful to Fernando Brand\~ao, Aram Harrow and Andreas Winter
for useful discussions. This work has been supported by the UK EPSRC's
``IRC QIP'', and the EU project QAP IST-3-015848.


\appendix

\section{Proofs of Lemmas 1 and 2}

In this appendix we describe the depolarization map $\Delta$ used in the proofs, and find some of its properties. We use the same notation as in the rest of the paper:
$\H=\bigotimes_{n}^N \H_n$, $\K=\bigotimes^N \CC^2$, and $\Omega$ is any separable completely-positive map whose input and output spaces become fixed by the context. We also show the following two lemmas, which are necessary for proving Theorems 1 and 2. \vs

{\bf Lemma 1.} {\em Let $\Omega$ be any separable map (\ref{pp}) transforming matrices acting on $\K$ to matrices acting on $\K$, and $\lambda\in [1/2,1)$, then
\begin{equation}\label{L11}
    [\Delta\circ\Omega\circ\Delta] (\lambda\id -\Phi)
    \leq \lambda\id -\Phi
\end{equation}
implies
\begin{equation}\label{L22}
    \ \ [\Delta\circ\Omega\circ\Delta] (\lambda\id-\Phi)
    = \lambda\id -\Phi \ .
\end{equation}}\vs

{\bf Lemma 2.} {\em Let $\Omega$ be any separable map (\ref{pp}) transforming matrices acting on $\K$ to matrices acting on $\H\otimes\K$, and $\lambda\in [1/2,1)$, then
\begin{equation}\label{L2}
    [(\id_\H \otimes \Delta_\K)\circ\Omega\circ\Delta] (\lambda\id -\Phi)
    = \sigma \otimes (\lambda\id -\Phi)
\end{equation}
implies that $\sigma$ is separable (\ref{fully-sep}).}\vs

\subsection{Notation}

Denote by ${\bf x}$ the $N$-bit string $x_1 x_2 \cdots x_N$ where $x_n\in\{0,1\}$, by
$\bar{\bf x}$ the $N$-bit string where each bit has the opposite
value as in ${\bf x}$, and by $\ket{\bf x}$ the vector
$\ket{x_1} \o\cdots \o\ket{x_N} \in\K$, where $\{\ket{0},\ket{1}\}$ is an orthonormal basis of $\CC^2$. The $N$-partite GHZ basis is defined
as
\begin{equation}\label{GHZbasis}
    \ket{\Phi_{\bf x}^\pm}=\frac{1}{\sqrt{2}}
    \left( \ket{\bf x} \pm \ket{\bar{\bf x}} \right)
    \quad \forall\, {\bf x} \mbox{ even ,}
\end{equation}
where ${\bf x}$ even means that $x_N=0$. There are $2^N$ such vectors and all of them are orthogonal.
(We impose $\x$ to be even in order not to doubly count states.) We
also define the unit-trace, positive semi-definite matrices
\begin{eqnarray}
    P_{\bf x} &=& \frac{1}{2}\left( \ket{\bf x}\!\bra{\bf x}
    + \ket{\bar{\bf x}}\!\bra{\bar{\bf x}} \right)
    \quad \forall\ {\bf x} \mbox{ even ,}\\
    P_\pm &=& \ket{\Phi_0^\pm}\!\bra{\Phi_0^\pm}\ .
\end{eqnarray}
In the next, the index $r$ takes values in the set
$\{+,-,2,4, 6,8,\ldots 2^N-2 \}$ and is used to label the states
$P_r$. For instance, one can write \mbox{$\id = 2\sum_r P_r - P_+ - P_-$}. \vs

\subsection{Depolarization protocol}

In \cite{dc} it is presented a useful LOCC protocol that transforms $N$-qubit states. The protocol is implemented through the following $(N+1)$ steps: (step 1) with probability 1/2 the $N$ parties apply $\bigotimes^N \!\sigma_x$ and with probability 1/2 they do nothing, \mbox{(step 2)} with probability 1/2 parties $\{1,N\}$ apply
$\sigma_z\otimes\sigma_z$ and with probability 1/2 they do nothing, \mbox{(step 3)} with probability 1/2 parties $\{2,N\}$ apply $\sigma_z\otimes\sigma_z$ and with probability 1/2 they do nothing, $\ldots$ (step N) with probability 1/2 parties $\{N-1,N\}$ apply $\sigma_z\otimes\sigma_z$ and with probability 1/2 they do nothing, \mbox{(step $N+1$)} the $N$ parties apply $\bigotimes_n^N \!\sigma_z^{\ y_n}$, where $y_2, y_3, \ldots y_N$ are independent (uniformly distributed) random bits, and $y_1 = \sum_{n=2}^{N} y_n \bmod 2$. Let $\Delta$ be the completely-positive map corresponding to this protocol. Because this protocol can be implemented by LOCC the map $\Delta$ is trace-preserving and separable. It is easy to check that for any $N$-qubit state $\varrho$ (acting on $\K$)
\begin{eqnarray}\label{depo}
    \Delta(\varrho) &=& \sum_r \varrho_r\, P_r\ , \\
    \varrho_\pm &=& \tr(P_\pm\, \varrho)\ , \\
    \varrho_\x &=& 2\, \tr(P_\x\, \varrho)\ \forall\, {\bf x} \mbox{ even .}
\end{eqnarray}\vs

\subsection{Characterization of $[\Delta \circ \Omega \circ \Delta]$-maps}

Let $\Omega$ be a separable completely-positive map that
transforms $N$-qubit states into $N$-qubit states. By the
Jamio{\l}kowski theorem \cite{J} we can write $\Omega(Z)
=\tr_\K(\Theta_{\K' \K} Z_\K\t)$ for any $Z$ acting on $\K$, where $\Theta_{\K' \K}$ is a
fully-separable $N$-partite state with two qubits per site. The $N$-qubit Hilbert space
$\K$ ($\K'$) corresponds to the input (output) of the channel $\Omega$. One can check that
\begin{equation}
    [\Delta\circ\Omega\circ\Delta](Z)=\tr_{\K}\!\Big(
    [\Delta_{\K'}\o\Delta_\K](\Theta_{\K' \K})\, Z_\K\t \Big)\ ,
\end{equation}
which, according to (\ref{depo}), implies that the state $\Theta$ associated to the map
$[\Delta\circ\Omega\circ\Delta]$ is of the form
\begin{eqnarray}\label{2qubit}
    &&\Theta = \sum_{r,r'} \Theta_{rr'} P_r\o P_{r'}\ ,\\
    \label{positiu}
    &&\Theta_{rr'} \geq 0 \quad \forall\ r,r'\ .
\end{eqnarray}
Because the maps $\Omega$ and $\Delta$ are fully-separable, the state $\Theta$ must be
PPT with respect to all bipartitions of the $N$ parties \cite{dc}.
Denote by $\varrho^{T_S}$ the matrix obtained when transposing in the basis $\{\ket{0}, \ket{1}\}$ of
the Hilbert spaces of the parties $S\subseteq \{1,\ldots N\}$
the matrix $\varrho$. One can check that for any
$S\subseteq\{1,\ldots N\}$
\begin{eqnarray}\label{t1}
    P_\pm^{T_S} &=& P_0 \pm \frac{1}{2}
    \left(\ket{\x_S}\!\bra{\bar{\x}_S}
    +\ket{\bar{\x}_S}\!\bra{\x_S} \right)\ ,
    \\ \label{t2}
    P_{\x}^{T_S} &=& P_{\x}\ ,
\end{eqnarray}
where $\x_S$ is the $N$-bit string with $x_n=1$ if $n\in S$ and
$x_n=0$ otherwise, for $n=1,\ldots N$. Without loss of generality
we assume that $S$ never contains the $N\th$ party, or
equivalently, the number with binary expansion $\x_S$ is
even. Using (\ref{t1},\ref{t2}) one can obtain the necessary and
sufficient condition for a state of the form (\ref{2qubit}) to be
PPT for all subsets of parties $S$, which is:
\begin{eqnarray}\label{ppt1}
  |\Theta_{++}-\Theta_{+-}+\Theta_{-+}-\Theta_{--}|
  -\Theta_{+{\bf x}}-\Theta_{-\x} &\leq& 0\hspace{10mm} \\
  |\Theta_{++}+\Theta_{+-}-\Theta_{-+}-\Theta_{--}|
  -\Theta_{{\bf x}+}-\Theta_{\x-} &\leq& 0 \\
  |\Theta_{+{\bf x}}-\Theta_{-{\bf x}}| -\Theta_{\x\x'} &\leq& 0 \\
  |\Theta_{{\bf x}+}-\Theta_{{\bf x}-}| -\Theta_{\x \x'} &\leq& 0 \\
  |\Theta_{+{\bf x}}-\Theta_{-{\bf x}}+\Theta_{\x+}-\Theta_{\x-}|
  \nonumber \\
  -\Theta_{++}+\Theta_{+-}+\Theta_{-+}-\Theta_{--} -\Theta_{\x \x} &\leq& 0 \\
  \label{ppt6}
  |\Theta_{+\x}-\Theta_{-\x}-\Theta_{\x+}+\Theta_{\x-}|
  \nonumber \\
  +\Theta_{++}-\Theta_{+-}-\Theta_{-+}+\Theta_{--} -\Theta_{\x \x} &\leq& 0
\end{eqnarray}
for all $\x,\x' \in \{2,4, 6,8, \ldots 2^N-2 \}$ with $\x'\neq\x$. \vs

\subsection{Proof of Lemma 1}

If we represent the Jamio{\l}kowski state associated to the map $[\Delta \circ \Omega \circ \Delta]$ as (\ref{2qubit}), equation  (\ref{L11}) is
equivalent to the following set of inequalities
\begin{eqnarray}\label{ku1}
    \lambda-1 -\lambda\sum_r \Theta_{+r} +\Theta_{++} &\geq& 0\ , \\
    \lambda -\lambda\sum_r \Theta_{-r} +\Theta_{-+} &\geq& 0\ , \\
    \label{ku3}
    2\lambda -\lambda\sum_r \Theta_{\x r} +\Theta_{\x +} &\geq& 0\ ,
\end{eqnarray}
for all $\x$ even. Now, for each $N\geq 2$ and $\lambda\in [1/2,1)$, the set
of inequalities (\ref{positiu},\ref{ppt1}-\ref{ku3}) defines a
linear-programming feasibility problem. Notice that each
inequality written with an absolute value is equivalent to two
plain linear inequalities. Fixing $N$ and $\lambda$ to some values, one can
obtain the polyhedron that contains all solutions (for example
with the software \cite{porta}). We have done this for different
values of $N$ and $\lambda$, obtaining always the polyhedron consisting of the single point
\begin{equation}\label{solution}
    \Theta\sol_{rr'}=\left\{
    \begin{array}{l}
        1\quad \mbox{if}\quad r=r'=\pm \\
        2\quad \mbox{if}\quad r=r'\in \{2,4, 6,\ldots 2^N-2 \} \\
        0\quad \mbox{otherwise} \\
    \end{array}\right. .
\end{equation}
This is precisely the Jamio{\l}kowski state corresponding to the depolarization
map $\Delta$:
\begin{equation}\label{}
    \Theta\sol = P_+\o P_+ +P_-\o P_- +2\sum_{\x} P_\x\o P_\x \ .
\end{equation}
In the next, we show that $\Theta\sol$ is indeed the unique
solution for any value of $\lambda$ and $N$.\vs

By exploiting the symmetry of equations \mbox{(\ref{positiu},
\ref{ppt1}-\ref{ku3})} one can get a system of 25 linear
inequalities and 10 unknowns, for any value of $N\geq 2$. Consider
the set of permutations $\{\pi\}$ that leave invariant the first
two elements in $(+,-,2,4, 6,\ldots 2^N-2 )$. If we perform the
transformation $\Theta_{rr'}\rightarrow \Theta_{\pi_r \pi_{r'}}$
the system of inequalities (\ref{positiu},\ref{ppt1}-\ref{ku3})
remains invariant. Therefore, if $\tilde{\Theta}_{rr'}$ is a
solution then also $\tilde{\Theta}_{\pi_r \pi_{r'}}$ is a
solution, for any permutation $\pi$ of the kind specified above.
Because the set of inequalities \mbox{(\ref{positiu},
\ref{ppt1}-\ref{ku3})} is linear, convex combinations of
solutions are also solutions. Hence, given a solution
$\tilde{\Theta}_{rr'}$ we can always generate another solution of
the form
\begin{equation}\label{}
    \Theta\inv_{rr'}=
    \frac{1}{(2^{N-1}-1)!} \sum_{\pi}
    \tilde{\Theta}_{\pi_r \pi_{r'}}\ ,
\end{equation}
where the sum is over the group of permutations specified above. It is clear that
the result of this average is invariant under transformations of
the form $\Theta\inv_{rr'}\rightarrow \Theta\inv_{\pi_r
\pi_{r'}}$. Therefore, the coefficients $\Theta\inv_{\x \pm}$ do
not depend on the value of $\x$, and analogously for
$\Theta\inv_{\pm \x}$. By applying group theoretic arguments \cite{C} one can see that the coefficients of the form
$\Theta\inv_{\x \x'}$ only depend on whether $\x=\x'$ or
$\x\neq\x'$. To see this, consider the representation of $\{\pi\}$ as orthogonal matrices which permute the components of vectors accordingly:
$O_{\x \x'}= \delta_{\x' \pi_{\x}}$. This representation decomposes into two irreducible ones: (i) the trivial representation spanned by the vector with all the entries equal to one ($v_\x=1\ \forall \x$), and (ii) the so called {\em standard} representation \cite{C}, corresponding to the complementary subspace. By Schur's Lemma \cite{C},
any matrix that commutes with all members of the group
$\{O_{\x\x'}\}$ has to be a linear combination of the projector
onto $v_\x$ and the identity. Hence, the restriction of
the matrix $\Theta\inv_{rr'}$ to $r,r'\in\{2,4,\ldots 2^N-2 \}$
can be expressed as $\Theta\inv_{\x\x'}=\alpha
+\beta\delta_{\x\x'}$, where $\alpha$ and $\beta$ are two numbers.
Concluding, each invariant state $\Theta\inv_{rr'}$ is completely
characterized by 10 real numbers: $\Theta_{++}, \Theta_{+-},
\Theta_{-+}, \Theta_{--}, \Theta_{+2}, \Theta_{-2}, \Theta_{2+},
\Theta_{2-}, \Theta_{22}$ and $\Theta_{24}$. (The choice of 2 and 4 as distinct even numbers is irrelevant.)\vs

Now, we can rewrite the set of inequalities
(\ref{positiu},\ref{ppt1}-\ref{ku3}) with only these 10 numbers:
\begin{eqnarray}\label{z0}
    \Theta_{++}, \Theta_{+-}, \Theta_{-+}, \Theta_{--} &\geq & 0
    \\
    \Theta_{+2}, \Theta_{-2}, \Theta_{2+}, \Theta_{2-}, \Theta_{22}, \Theta_{24}
    &\geq & 0
  \label{b0}
\end{eqnarray}
\begin{eqnarray}
  |\Theta_{++}-\Theta_{+-}+\Theta_{-+}-\Theta_{--}|
  -\Theta_{+2}-\Theta_{-2} &\leq 0& \hspace{8mm} \\
  |\Theta_{++}+\Theta_{+-}-\Theta_{-+}-\Theta_{--}|
  -\Theta_{2+}-\Theta_{2-} &\leq 0& \quad \\
  |\Theta_{+2}-\Theta_{-2}| -\Theta_{24} &\leq 0&  \\
  |\Theta_{2+}-\Theta_{2-}| -\Theta_{24} &\leq 0& \\
  |\Theta_{+2}-\Theta_{-2}+\Theta_{2+}-\Theta_{2-}| +
  \\ \nonumber
  -\Theta_{++}+\Theta_{+-}+\Theta_{-+}-\Theta_{--} -\Theta_{22} &\leq 0& \\
  \label{ppt6}
  |\Theta_{+2}-\Theta_{-2}-\Theta_{2+}+\Theta_{2-}| +
  \\ \nonumber
   +\Theta_{++}-\Theta_{+-}-\Theta_{-+}+\Theta_{--} -\Theta_{22} &\leq 0&
\end{eqnarray}
\begin{eqnarray}
    \lambda\left(\Theta_{++}+\Theta_{+-}+g_N\Theta_{+2}\right)
    -\Theta_{++} -1+\lambda \hspace{6mm}
    \label{b1} \\ \nonumber \leq 0 \hspace{6mm} \\
    \lambda\left(\Theta_{-+}+\Theta_{--}+g_N\Theta_{-2} \right)
    -\Theta_{-+} -\lambda \hspace{6mm}
    \\ \nonumber \leq 0 \hspace{6mm} \\
    \label{z100}
    \lambda\left(\Theta_{2+}+\Theta_{2-}+(g_N-1)\Theta_{24}+\Theta_{22}\right)
    -\Theta_{2+} -2\lambda \hspace{6mm}
    \\ \nonumber \leq 0 \hspace{6mm}
\end{eqnarray}
For any $N$, this system has 10 unknowns, but the coefficient
$g_N=2^{N-1}-1$ still depends on $N$. By direct substitution one
can check that the point $\Theta\sol$ in (\ref{solution}) is a
solution of this system. Let us prove that it is unique. If a new linear
inequality can be expressed as a linear combination with
non-negative coefficients of the inequalities
(\ref{z0}-\ref{z100}), then it must be satisfied by all the
solutions of this system. With the help of a computer, we have
expressed each of the following 20 inequalities
\begin{eqnarray}
    \nonumber
    1 &\leq& \Theta_{++},  \Theta_{--} \leq 1 \\
    \nonumber
    2 &\leq& \Theta_{22} \leq 2 \\
    \nonumber
    0 &\leq& \Theta_{+-}, \Theta_{-+}, \Theta_{+2}, \Theta_{-2},
    \Theta_{2+}, \Theta_{2-}, \Theta_{24} \leq 0\
\end{eqnarray}
as a linear combination with non-negative coefficients of the ones
in (\ref{z0}-\ref{z100}), with the coefficients explicitly
depending on $g_N$ and $\lambda$. These coefficients are well
defined for any $N\geq2$ and $\lambda\in(1/2,1)$, but
unfortunately, some of them become singular when $\lambda=1/2$.
Later we analyze this case. This implies that, when $\lambda\neq
1/2$, all the solutions of (\ref{z0}-\ref{z100}) satisfy these
inequalities, but it is straightforward to see that the only point
satisfying them is $\Theta\sol$, as we wanted to prove.\vs

For $\lambda=1/2$ and $N=2$ we have solved the polyhedron
(\ref{z0}-\ref{z100}) with the software \cite{porta}. In this case
the set of solutions form a convex cone with vertex in
$\Theta\sol$. The software \cite{porta} also outputs the
generators of the cone, which can be written as $\Theta^{[r]} =
P_r\o(P_+ +P_-)$, for all $r$. Then, all the solutions are of the
form $\Theta^{\mbox{\scriptsize cone}} = \Theta\sol +\sum_r\tau_r \Theta^{[r]}$ with $\tau_r\geq 0$.
Because $\Theta_{+2}, \Theta_{-2}, \Theta_{24}, g_N, \lambda \geq 0$, and $g_N$ increases with $N$, all solutions of (\ref{z0}-\ref{z100}) for any $N> 2$ are contained in the set of solutions for $N=2$, denoted $\{\Theta^{\mbox{\scriptsize cone}}\}$.
All points in $\{\Theta^{\mbox{\scriptsize cone}}\}$ have the property that $\Theta_{+2}= \Theta_{-2}=
\Theta_{24}= 0$, therefore, increasing the value of $g_N$ does not
reduce the set of solutions. Thus, for any $N$, the set of solutions of (\ref{z0}-\ref{z100}) is precisely $\{\Theta^{\mbox{\scriptsize cone}}\}$. When  $\lambda=1/2$ we have
that $\tr[(P_+ + P_-)(\lambda\id -P_+)]=0$, then all the maps in $\{\Theta^{\mbox{\scriptsize cone}}\}$ give the output $(\lambda\id- P_+)$ when the input is $(\lambda\id- P_+)$. Concluding, any $\Omega$ for which (\ref{L11}) holds is such that (\ref{L22}) must hold too. $\Box$\vs

\subsection{Proof of Lemma 2}

Applying Schur's Lemma \cite{C} on the outer depolarization map $\Delta_\K$,
if $\Omega$ maps states acting on $\K$ to states acting on
$\H\o\K$ then
\begin{eqnarray}\label{f}
    [(\id_\H\o\Delta_\K)\circ\Omega\circ\Delta] (\lambda\id-\Phi)=
    \\ \nonumber
    (\lambda-1)\, \sigma_+\o P_+ +\lambda \sum_{r\neq +} \sigma_r \o
    P_r\ ,
\end{eqnarray}
where $\sigma_r$ are unit-trace, positive semi-definite matrices acting on $\H$. Because the maps
$\Delta, \Omega$ and the states $P_\x$ are fully-separable, the states
\mbox{$\sigma_\x = \tr_{\K}[(\id_\H\o\Delta_\K)\circ\Omega\circ\Delta] (P_\x)$} must be fully-separable too, for $\x\in\{2,4,\ldots
2^N-2\}$. The matrix $\sigma \o (\lambda\id-P_+)$ is product, and according to (\ref{L2}) is equal to the right-hand side of (\ref{f}). This, together with the fact that the matrices $P_r$ are orthogonal (with respect to the Hilbert-Schmidt scalar product), implies that all the matrices $\sigma_r$ must be
equal. Then $\sigma=\sigma_\x$, which is
fully-separable. $\Box$

\end{document}